# THE ADAPTATION OF COMPLEXITY IN THE EVOLUTION OF MACROMOLECULES

Nilou Ataie

nataie@queens.org

47-754 Kamehameha Hwy, Kaneohe HI 96744

**ABSTRACT**

Enzymes are on the front lines of evolution. All living organisms rely on highly efficient, specific enzymes for growth, sustenance, and reproduction; and many diseases are a consequence of a mutation on an enzyme that affects its catalytic function. It follows that the function of an enzyme affects the fitness of an organism, but just as rightfully true, the function of an enzyme affects the fitness of itself. Understanding how the complexity of enzyme structure relates to its essential function will unveil the fundamental mechanisms of evolution, and, perhaps, shed light on strategies used by ancient replicators. This paper presents evidence that supports the hypothesis that enzymes, and proteins in general, are the manifestation of the coevolution of two opposing forces. The synthesis of enzyme architecture, stability, function, evolutionary relationships, and evolvability shows that the complexity of macromolecules is a consequence of the function it provides.

## WHAT IS LIFE?

Although the question "what is life?" has been argued by thousands of the worlds brightest minds, no agreement has been reached on the exact definition because many of the distinctions of what constitute life are arbitrary. Observations show that a major difference between living and most non-living systems is that living systems have the ability to manifest, maintain, and replicate seemingly highly improbable distributions of matter and energy. For instance, a snapshot of any organism at any given time reveals a conglomeration of molecules and ions that not only display patterns in their arrangement in the space provided by the organism, but also display patterns in the atoms that make up the large molecules. A snapshot of the organism at a later time reveals that although the *patterns* of atoms in space have not changed much, the system has moved toward exchanging the individual atoms. For example, the atoms that make up a human being are not the same atoms that formed that human ten years earlier. The matter has exchanged with the environment, but that human still exists. Thus, organisms have the ability to write, store, and pass the information needed to unfold and maintain complex patterns of matter in space through time. Life is a stable, yet dynamic, adaptive chemical system – a mass of ever-exchanging atoms that exhibit specified complexity (Orgel 1973, Davies 1999) - the question is: how does this happen? How do molecules replicate, adapt, change, breed complexity? What is the origin of heredity?

Heredity could have begun as a chance initiation of an autocatalytic system - a system in which the products of catalysis are used to catalyze the formation of more products (Kauffman 1993, Dawkins 2004). Our most ancient ancestors, the first replicators, may have been small molecule catalysts - evolving, adapting complexity, and by doing so, "living", existing in harmony with physical entropy. The question becomes: what are the physical forces that support molecular replication and growth, and how are these forces acquired? We may only be able to speculate on how early molecules evolved because today's earth is very different from the young, volatile earth that life first emerged from, but we can examine how the catalysts of today evolve, and by doing so, we may gain insight into the forces that support molecular evolution in general.

Enzymes are catalysts that evolve, and without them, there would be no life. They are composed of thousands of atoms that are bound together in very specific ways, making them some of the largest, most complex molecules in the known universe. They are also some of the most powerful catalysts in the world. The fundamental mechanisms of evolution lie hidden in the relationship between enzyme structure and function.

**PROTEINS ARE COMPLEX MOLECULES THAT OBEY THE LAW OF PHYSICAL ENTROPY**

According to the second law of thermodynamics (a law based on observations of how matter and energy behave in time), matter and energy are spontaneously driven toward a state of higher disorder – a state that best flattens the probability distribution of matter and energy in the space available through time. The patterns observed in the distribution of the atoms in complex, specifiable molecules like proteins create the illusion that they exist in defiance of the second law. However, if we follow the fate of the protein atoms we would see that the second law is never violated; the atoms organized at *time 1* had indeed dissipated to a state of greater disorder. The complex molecules that make up living organisms do not defy the law of entropy at all - so how do they exist in the first place?

This paradox is revealed when it is realized that all proteins perform a function, and the function is a direct consequence of their structure, which is a direct consequence of the genetic code. The function is almost always essential to the survival and proliferation of the organism, and hence, the replication of the molecule. A necessary condition is that the rate at which the information (to replicate) is passed (the function is performed) is greater than the rate of the breakdown and dissipation of the molecule; because thermodynamics says nothing about rates, the second law is never violated.

To give an example, at any given moment in time human beings have numerous molecules of the enzyme carbonic anhydrase diligently working to maintain the essential blood pH. Each molecule of carbonic anhydrase will break down, dissipate, succumb to entropy – but at a rate slower than the rate it takes the enzyme to catalyze the reactions necessary for the organism to survive (due to the sheer number of interacting components, organisms are quasi-adiabatic, but they are, ultimately, open systems). In the case of carbonic anhydrase, the function performed gets the enzyme replicated in the same organism, and ultimately, in the progeny of that organism. Random mutations occur, but if the mutations affect the catalytic function of the enzyme in a way that has adverse consequences for the organism - the organism fails, and that variant of the enzyme does not replicate, and hence, does not exist. Enzyme sequences are selected for, and the reason that specified, complex molecules exist is because they can do the function that transfers the information necessary for them to exist. Seen through the lens of natural selection, matter can organize and localize in a universe where the second law is never violated. How does matter spontaneously organize to form large structures that display symmetry and organization? By an increase in entropy of course.

**STRUCTURE AND SYMMETRY CAN MANIFEST BY AN *INCREASE* IN PHYSICAL ENTROPY.**

The law of physical entropy states that in a given system in a time, Δt, the system spontaneously moves toward a uniform probability distribution of matter and energy in the space available. This simply means that the system spontaneously unfolds, if not already unfolded, so that *on average* the probability of finding atom *x* in space *x* at time *2* has moved in the direction of matching the probability of finding atom *x* in space *y* at time *2* (for a review see Atkins 2007 and Ben-Naim 2008). However, the available space is limited by restraints imposed by the system, and because of the restraints, patterns in the organization and compilation of matter may manifest as the entropy of the system increases. Because the second law applies to the average behavior of the particles, and nuclear, electrostatic, and mechanical forces limit space, systems moving in the direction of greater physical entropy can manifest large-scale structure and symmetry.

To give an example, if the detergent *n*-Dodecyl β-D-maltoside (DDM) is mixed with water to give a final concentration of 1 mM (at room temperature and pressure), emergent structure will *spontaneously* manifest in countless near perfect spheres called micelles where dozens of detergent molecules amass and organize in space. Although the physical entropy of the system moves toward maximizing disorder, emergent structure and symmetry are formed. Why? Optimal flattening of the probability distribution of the components of this particular system means micelle formation. Repulsive electrostatic interactions restrain the conformation of the water molecules around the hydrophobic moiety of the detergent, and the surface area in which the water molecules must order themselves is reduced as a consequence of micelle formation. Although the detergent molecules are more restrained in a micelle, the maximum entropy of the system (the flattest probability distribution) is achieved when micelles form (figure 1).

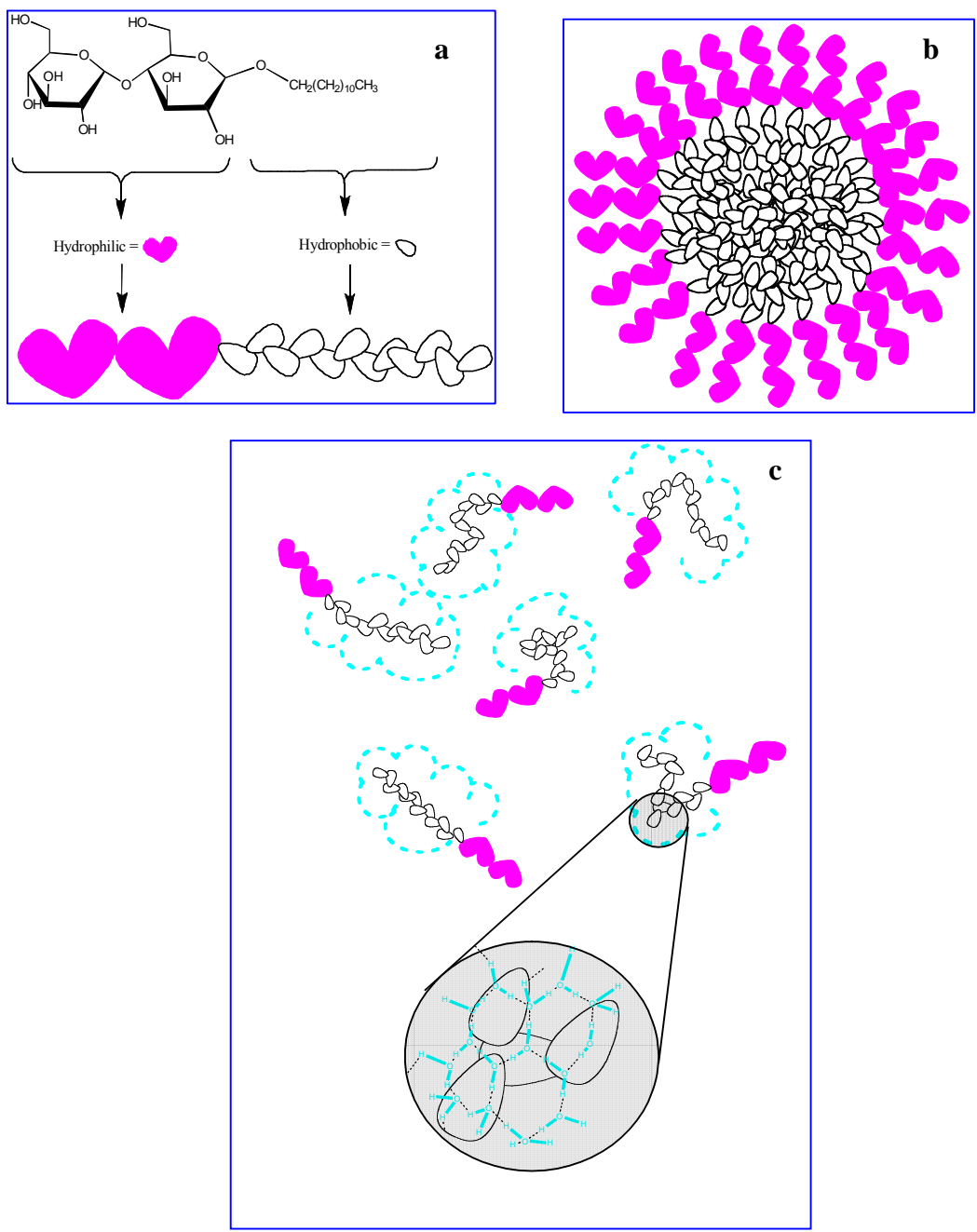

**Figure 1**: **Structure and Organization May *Spontaneously* Manifest as a Consequence Of An Increase in Entropy**. When the amphiphilic molecule DDM, **a**, is mixed with water to give a final concentration of 1 mM (at room temperature and pressure), large scale structure and symmetry will *spontaneously* manifest in countless near perfect spheres called micelles, **b**. The water molecules around the hydrophobic moiety of the detergent must order themselves**, c,** and this is greatly reduced as a consequence of micelle formation. Although the detergent molecules are more restrained in a micelle, the maximum entropy of the *system* is achieved when micelles form.

Upon construction, most of nature's proteins spontaneously, rapidly, and faithfully fold into a very particular form; and just like micelles, their form is driven by physical entropy and molded by electrostatics and mechanics. Unlike the circumstances that lead to micelle formation, the circumstances that lead to the formation of proteins in biological systems are set-up by natural, physical processes and not an intelligent scientist. Unlike the micelles, proteins are functioning agents in a vast interconnected autocatalytic network of complex molecules. Through the synthesis of protein structure, function, stability, evolutionary relationships, and evolvability this paper shows that the specified complexity of macromolecules is simply a consequence of the function it provides.

**ENZYME ARCHITECTURE**

Enzymes are large molecules constructed by the linkage of smaller, more stable molecules. The order in which the subunits are linked together determines the structure of the enzyme, and hence, the enzymatic function. Most enzymes are proteins, whose structure-function springs to existence from a particular sequence of twenty chemically unique amino acids. Once linked together, the forces of attraction and repulsion mediate to form an energetically stable protein fold. The order of linkage and the chemical properties of the amino acids predetermine the architecture of the protein (figure 2)

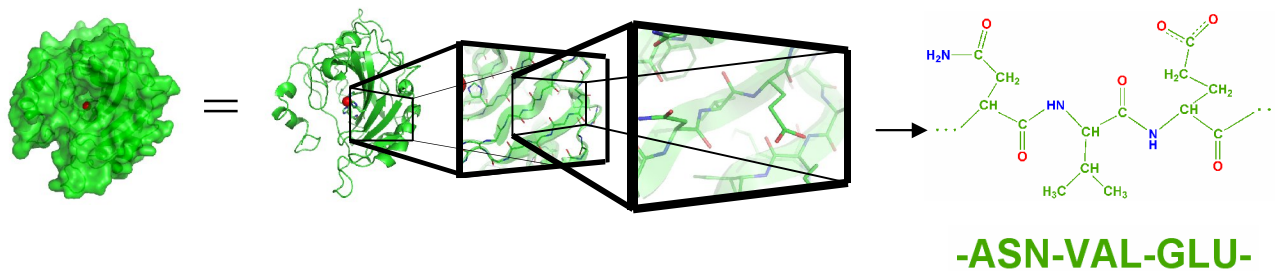

**Figure 2: Proteins are Complex Molecules: They are Composed of Thousands of Atoms Organized and Localized in Space.** Most enzymes are proteins, macromolecules composed of thousands of atoms organized and localized in space. Twenty chemically unique amino acids define the building blocks of all proteins. The sequence of the amino acids predetermines the physical nature of the protein. Although the sequence variations are near infinite, relatively few variations are represented in proteins of known sequence and structure. The physical nature of a sequence of amino acids is determined by a mediation of many forces, and only in a small percentage of sequences does this mediation produce a folded protein. Figure made with PyMOL.

The twenty amino acids encompass a range of chemical properties. Although the head groups are all the same, the side chains, which define the amino acids, are all chemically different. They can be polar, nonpolar, negatively charged, positively charged, short, long, bulky, flat, easily ionized, less easily ionized, hydrogen bond accepting, hydrogen bond donating, and both hydrogen bond accepting and donating (for a review of protein structure see Petsko & Ringe 2004). With this broad array of starting materials and their infinite number of arrangements the number of sequence variations and folds would seem endless. But this is not what is observed. Decades of data collection show that relatively few arrangements of amino acids are represented in proteins of known sequence and structure. For example, many genes code for proteins that harbor the sequence patterns N-P-N-P-N-P-N-P-N-P and P-N-N-P-N-N-P-P-N-N-P-P, where N = a non-polar residue and P = a polar residue. Nearly all of the tens of thousands of proteins of known structure contain specific structural motifs called alpha helices and beta sheets. Beta sheets are often composed of amino acid sequences that reflect the first pattern described above, and alpha helices are often composed of sequences that reflect the second pattern (figure 3). In addition, those patterns are often only found in their respective structural motifs. Because of this strict observed association it is often possible to predict alpha helices and beta sheets from sequence information alone, indeed, this method is used in protein structure prediction (Bowie et al. 1991, Zhang & Skolnick 2005). Mutations are random, but protein sequences are not randomly distributed. Known sequences sample only a small portion of sequence space, known structures, a small portion of configurational space. Out of all the possible sequences and structures that can be imagined, only a few are most probably observed. Thus, it can be said that proteins exhibit low "informational entropy" (Dewey 1996, Strait & Dewey 1996). The question is - how?

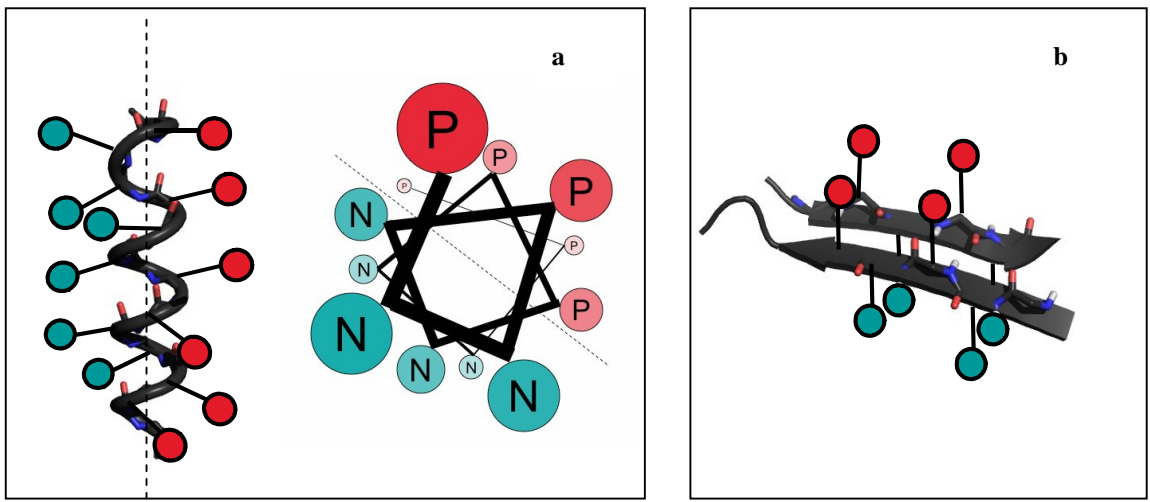

**Figure 3: Nearly All of The Tens of Thousands of Proteins of Known Structure Contain Specific Structural Motifs Called Alpha Helices And Beta Sheets – Both of Which Are Often Associated with Sequence Patterns.** Decades of data collection show that many genes code for proteins that harbor the sequence patterns N-P-N-P-N-P-N-P-N-P and P-N-N-P-N-N-P-P-N-N-P-P, where N = a non-polar residue and P = a polar residue. Beta sheets, **b**, are often composed of amino acid sequences that reflect the first pattern, and alpha helices, **a**, are often composed of sequences that reflect the second pattern. In addition, these patterns are often only found in their respective structural motifs. These types of associations make it possible to predict structural motifs from sequence information alone. Figure made with PyMOL.

It is important to remember that all proteins serve a function and that function has a beneficial effect on the survival and replication of the organism, and hence, the replication of the gene coding for the protein. The function is a direct consequence of the structure, which is a direct consequence of the genetic code. If the function has no beneficial effect on the survival and replication of the organism, when random mutations change the sequence coding for that protein, the organism may not be affected, so in this way too, the protein will soon cease to exist. Thus, most of the proteins that do exist – the observed patterns in sequence and structure - satisfy some functional *need*, because if they did not they would not exist. The observed gene sequences are selected for, so how do these patterns benefit function?

To perform its function a protein must take a very particular form. The protein must fold so that the reactive groups are physically and chemically primed to perform their function. Studies show that the physical nature of a sequence of amino acids is determined by a mediation of several forces, and only under certain circumstances does this mediation result in a folded protein. The driving force of folding is the hydrophobic effect, but many forces determine if the fold is stable enough to exist. Steric strain, hydrogen bond satisfaction, and ionic pairing are all essential to the stability of the folded state. The latter forces do not encourage folding, however, if left unsatisfied, they may discourage it (Fersht 1985, Fersht et al. 1987, Milla et al. 1994, Wang et al. 2006, Waldburger et al. 1995).

The unfolded protein has much higher conformational entropy than the folded protein – there are a lot more ways of being unfolded than folded. But just like the micelle experiment described earlier, the conformational entropy of water must also be considered, and in the unfolded state many water molecules can do nothing but form an ordered cage around the exposed hydrophobic residues. The surface area in which the water molecules must order themselves in greatly reduced in the folded protein, where most of the hydrophobic residues are buried in the protein core (figure 4).

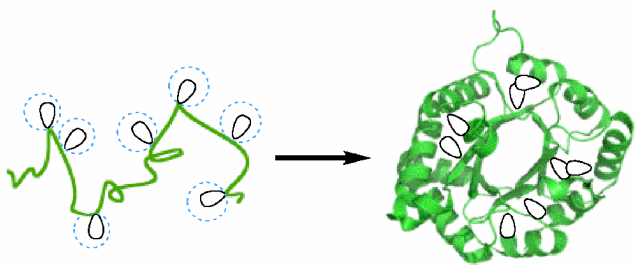

**Figure 4: The Hydrophobic Effect Drives Protein Folding.** In the unfolded state water molecules can do nothing but form an ordered cage around the exposed hydrophobic residues. The surface area in which the water molecules must order themselves is greatly reduced in the folded protein, where most of the hydrophobic residues are buried in the protein core. The energetic contribution this phenomenon makes to folding is called the hydrophobic effect (Kellis et al. 1989, Dill 1990, Serrano et al. 1991, Weihua et al. 2005). Figure made with PyMOL.

Hydrogen bonds and ionic interactions do not drive folding because they are satisfied in the unfolded state. However, if favorable charge interactions are not satisfied in the folded state, the fold becomes less stable (figure 5). Mutational studies show that when hydrophobic residues replace buried salt-bridges proteins remain active and have *greater* stabilities (Waldburger et al. 1995). Nearly every soluble, nature-made, functional protein of known structure has a hydrophobic core, and polar substitutions of hydrophobic core residues destabilize folds. Thus, observations show that *in general* the hydrophobic effect drives folding and electronic and steric strain destabilize the fold. This is a general observation and there are exceptions to this rule. Metal ions may stabilize and energize folds, for example, the zinc ion has a significantly greater affinity for the protein-zinc-finger-motif than it has for water. This energy disparity is used to pay the cost of holding the functional folded state. In this extreme example folding is driven mainly by electrostatics with a small contribution by the hydrophobic effect (Cox & McLendon 2000).

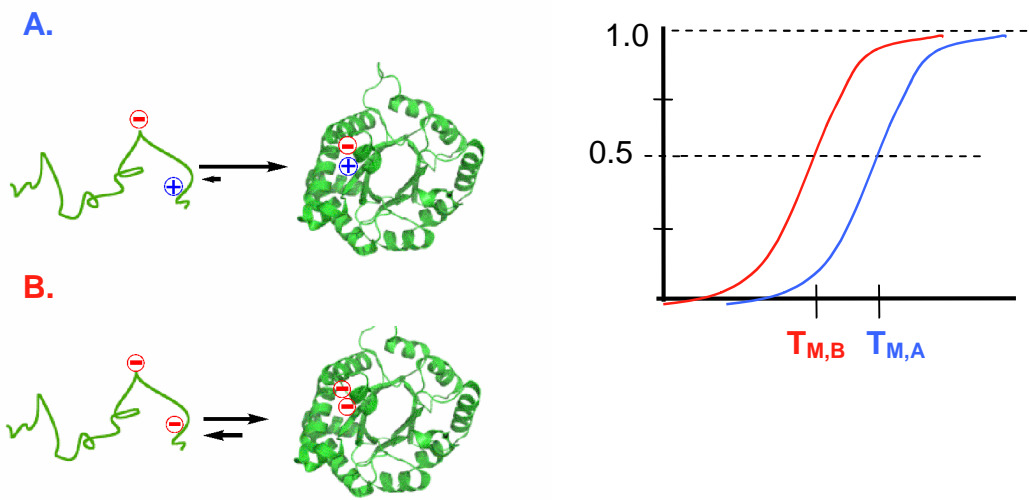

**Figure 5: Unsatisfied or Unfavorable Charge Interactions Affect the Stability of the Folded Protein.** Hydrogen bonds and ionic interactions do not drive folding because they are satisfied in the unfolded state. However, if favorable charge interactions are not satisfied in the folded state the fold becomes less stable (Fersht 1985, Fersht et al. 1987, Milla et al. 1994, Wang et al. 2006). For example, if a paired ionic interaction within the core of fold A. is broken by the removal of one charge, or the replacement of one charge with the opposite charge, the stability of the fold is lowered (fold B.). The stability, or strength, of a fold may be equated to the probability of the unfolded state as a function of temperature (graph). The blue curve represents the probability of fold A. being unfolded as a function of temperature, the red curve represent the same for fold B. The temperature at which the probability of the unfolded state is 50 % ($T_M$) is greater for fold A. than for fold B. Figure made with PyMOL.

In light of the forces that drive and stabilize protein folds, the observed sequence patterns in the genetic code and the associated three-dimensional structures turn out to be a good way of designing a protein scaffold. First, the periodicity of helices and sheets allows them to form surfaces that are polar on one side and nonpolar on the other (figure 3). This provides a means to enhance the amplitude of the hydrophobic effect and maintain polar interactions with water. Second, helices and sheets can accommodate well-satisfied main-chain hydrogen bond interactions, thereby minimizing unfavorable charge interactions. And third, all of this is done with little to no mechanical strain between adjacent side chains as helices and sheets can accommodate the preferred torsional angle of the side-chains (figure 6). It is tempting to speculate that the reason why proteins are chiral is because chirality offers the construction of helices and sheets.

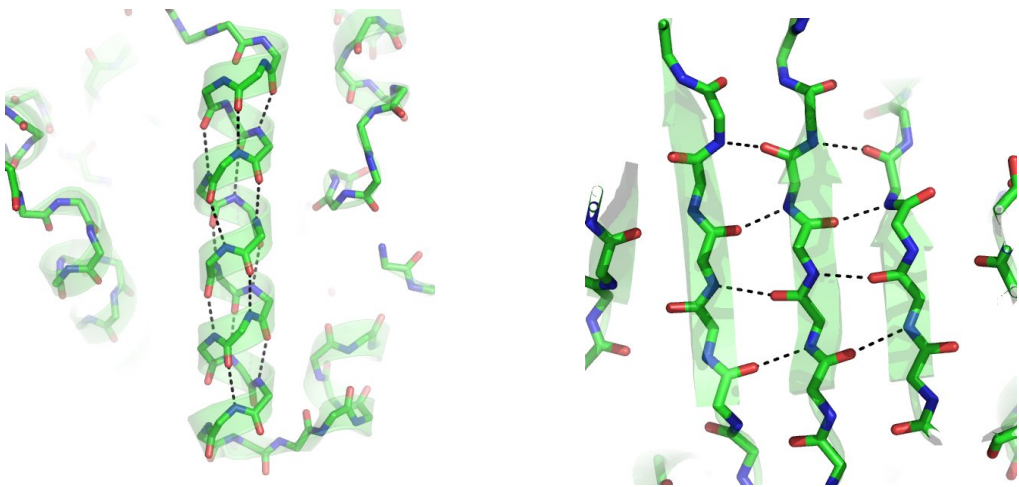

**Figure 6: Alpha Helices and Beta Sheets Provide Support to the Folded State**. Specific structural motifs called alpha helices and beta sheets support the folded state in a number of ways. First, the periodicity of helices and sheets allows them to form surfaces that are polar on one side and nonpolar on the other. This provides a means to enhance the amplitude of the hydrophobic effect and maintain polar interactions with water. Second, helices and sheets accommodate well-satisfied main chain hydrogen bonds, thereby minimizing unfavorable charge interactions. And third, all of this is done with little to no mechanical strain between adjacent side chains, as helices and sheets can accommodate the preferred torsional angles of the side chains (For a review see Petsko and Ringe 2004). Figure made with PyMOL.

The reason some sequences are more probably observed than others is because those sequences have what it takes to form three-dimensional functional proteins. And three-dimensional functional proteins are needed to replicate the code. Alpha helices and beta sheets are a good way of designing a protein scaffold and, indeed, it is hard to find a protein without them. The most common fold, the TIM Barrel, is almost completely composed of alpha helices and beta sheets and it is found in numerous non-homologous enzymes, of varying functions - leaving it unclear as to if they are divergently or convergently related (Farber & Petsko 1990, Reardon & Farber 1995, Wierenga 2001, Nagano et al. 2002). What is crystal clear is that when it comes to matters of catalytic function, the TIM Barrel scaffold has a selective edge.

**THE ORIGIN OF ENZYMATIC RATE ENHANCEMENT**

Like all catalysts, enzymes increase the rates of chemical reactions without being consumed. Reactions that would normally take hours, days, or years to complete may be finished in fractions of a second with the assistance of the right enzyme. Despite having thermodynamically favored product formation, reactions may take a long time to complete because the transition states the reactants must go through are energetically unstable, and hence, rarely manifest. The enzyme helps speed the reaction by providing a space where the transitioning species is less unstable, and hence, can more frequently materialize (figure 7).

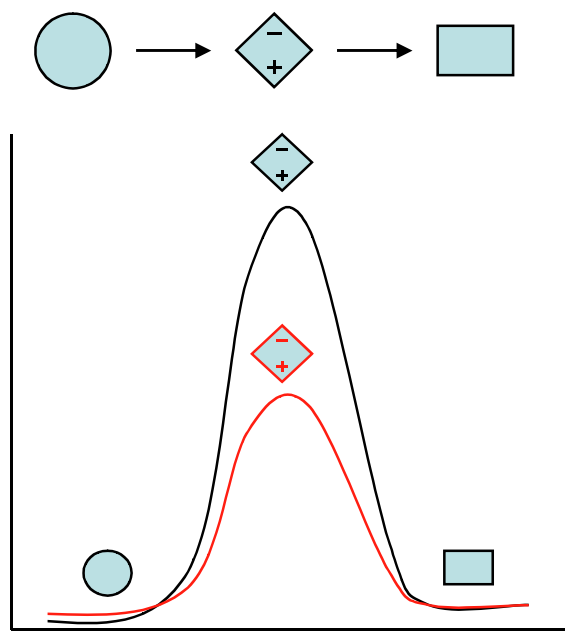

**Figure 7: Enzymes Provide a Space Where the Transitioning Species Can More Readily Manifest.** In this simple model, the rate at which the substrate, represented as a blue circle, turns to product, represented as a blue rectangle, depends on how easy it is to form the transition species (blue diamond), which exhibits charge separation. Despite having thermodynamically favored product formation, reactions may take a long time to complete because the transition states the reactants must go through are energetically unstable, and hence, rarely manifest. The enzyme helps speed the reaction by providing a space where the transitioning species is less unstable, and hence, can more frequently materialize.

Life processes depend on highly efficient, specific catalysts, and enzymes happen to be some of the most powerful in the world. Besides harboring complex patterns, enzymes are also incredibly large molecules, usually made-up of hundreds of amino acids – thousands of atoms. The catalytic centers of enzymes, the active sites, consist of only a handful of amino acids, where relatively few protein atoms ever come in contact with the reactants (Figure 8). It appears as if the production of the rest of the enzyme is a colossal waste of energy - but evolution is not so wasteful. Organisms expend a great deal of energy and resources constructing large proteins and any portion of a protein that does not aid in function is subject to deletion by natural selection: if organisms could get away with producing smaller proteins they would – or at least *some* would - but all enzymes are large, and all enzymes are complex. Thus, when it comes to matters of enzymatic function there must be a *need* to be a large complex molecule.

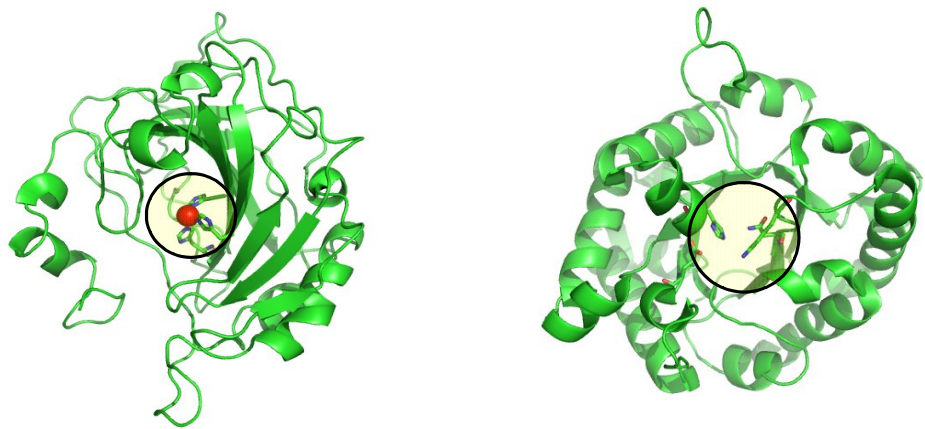

**Figure 8: Active Sites Are Composed of A Small Portion of the Enzyme, Where Relatively Few Residues Come in Contact With the Reactants**. As is typical of most enzymes, the active site of human carbonic anhydrase II, left, and trios phosphate isomerase, right, consist of a small localized region on the enzymes, where relatively few of the protein atoms come in contact with the reactants. It is important to remember that evolution tends to be a good engineer of structures optimized toward function – and organisms expend a great deal of energy and resources constructing large proteins. Thus, a theory of enzymatic rate enhancement must account for the *entire* molecule. Figure made with PyMOL.

Notwithstanding the need to understand how enzymes work, the forces at work in enzyme active sites are not readily available for observation and quantification, making the study of the origin of enzymatic rate enhancement difficult and the subject of many debates. Over the years several theories have been proposed as to how enzymes enhance the rate of chemical reactions, but only a few of these theories include an evolutionary perspective by addressing a clear relationship between structure and function. Two prominent theories that *do* account for the structure-function relationship are the preorganization theory and the entatic theory (Warshel 1978, Warshel et al. 1988, Warshel 1998, Vallee & Williams 1968, Williams 1972). These theories share the viewpoint that strain, near or at the active site, is a consequence of features that enhance catalysis, and that the rest of the enzyme has evolved stabilizing interactions to pay for this strain. Accordingly, complexity and function coevolve, exerting selective pressure on one another. Conceptually, any fold that can accommodate the functional strain will do. Some of the ever-increasing number of studies that support the strength-strain perspective will be presented and discussed shortly.

The preorganization theory is a powerfully predictive theory that has stood the test of time. When the enzymatic reaction is compared to a reaction in water that uses the same mechanism, chemical and computational results show that the catalytic power of enzymes originates in electrostatic stabilization provided by the preorganization of complementary dipoles around functional charges and the charged transition states (Warshel 1978). Figure 9 illustrates a simple example: in the space surrounded by water molecules, the transitioning species is accommodated when the water molecules are oriented in a rare, energetically unfavorable conformation that complements the charge on the transitioning species. In the space provided by the enzyme, the transitioning species is accommodated by complementary charges that are preoriented and stabilized by the fold – which has been optimized by evolution to accommodate the mechanical and electronic strain that is produced as a consequence of preorganization (Warshel et al. 1988, Warshel 1998). In short, complex structural features can do the work necessary to set up the space the substrate needs to transition into a product. It is important to note that for the purpose of clarity, this paper presents a very simple view of catalysis. Enzymes do not function by binding the transitioning species alone, they also function by destabilizing the binding of other species. For an enzyme that needs to turnover quickly, it may be essential that the binding affinities of products and other ligands are low. Preorganized features can also be used to destabilize the binding of unfavorable ligands.

The entatic theory, or the direct use of strain in rate enhancement, was first postulated as a result of observations made of the aberrant properties of metal ions in metalloenzyme active sites (Vallee & Williams 1968). When complexed to proteins, the metal ions of metalloenzymes often exhibit unusual, distorted coordination geometries and high reactivities. Chemical reactions catalyzed by metal ions often require the metal ions to change coordination geometries. Based on these observations and others, the entatic theory postulates that by holding the metal ion in a geometrically strained conformation the protein fold helps facilitate coordination geometry change, and hence, the rate of the reaction. Like the preorganization theory, the entatic theory relates enzymatic rate enhancement to strain paid for by the folding energy; unlike the preorganization theory, the entatic theory is not clear on the definition of strain (which can be mechanical or electronic) and the energetic contribution of that strain. The preorganization theory and the entatic theory are not mutually exclusive - coordination distortion may be a phenotype of the fine-tuning of the electrostatic effect, for instance, coordination distortion may polarize the charge/dipole complex and thereby enhance the electrostatic effect.

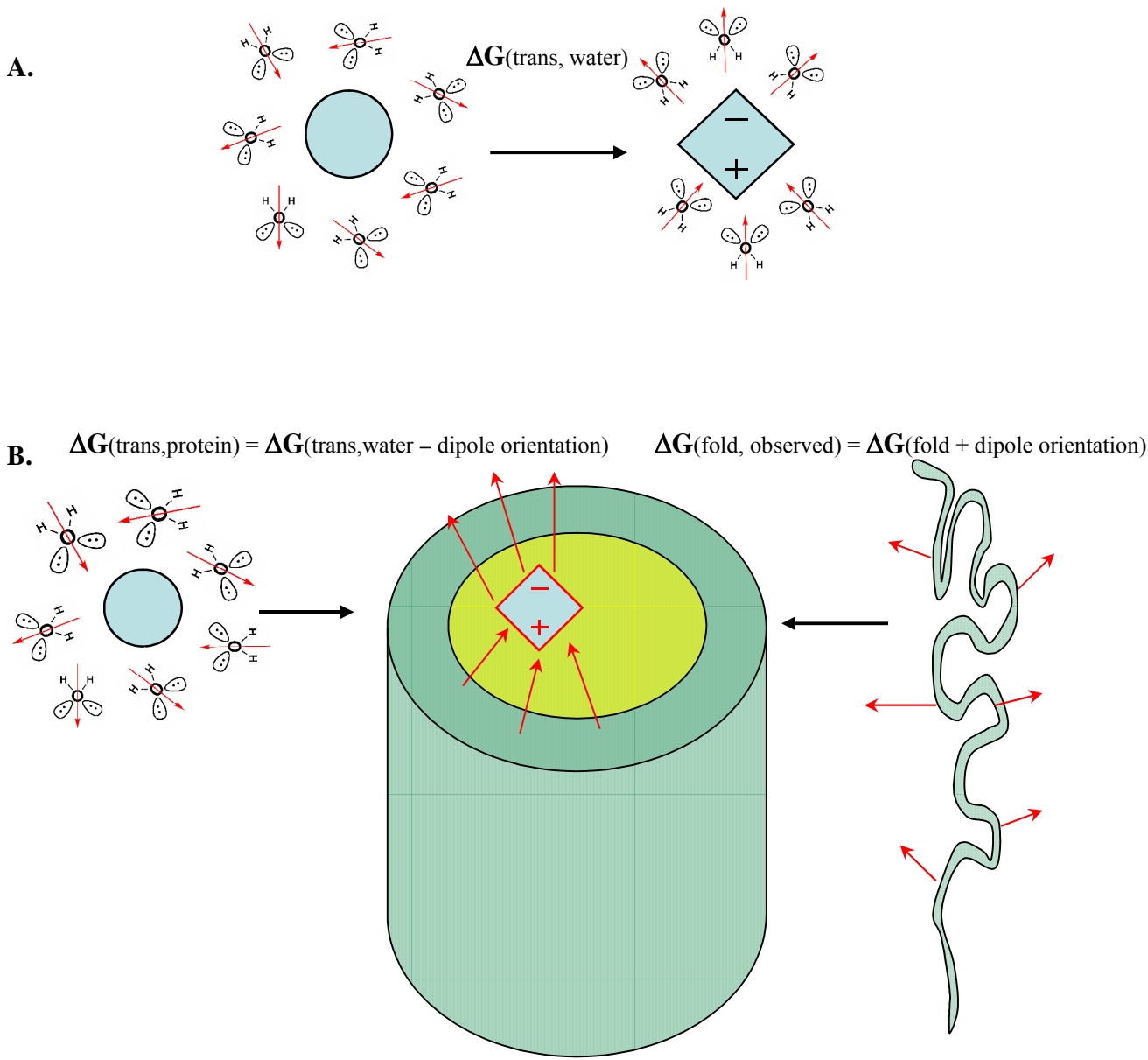

**Figure 9: Preorganized Dipoles Increase the Rate of a Reaction By Providing A Space Where the Transitioning Species is More Stable. The Electronic and Steric Strain Produced by Preorganization is Paid for By Folding Energy.** In the space surrounded by water molecules, A., the probability of the manifestation of the transitioning species is increased as the water molecules orient in a rare conformation, which is complementary to the transitioning charge. In the space provided by the enzyme, B., the probability of the manifestation of the transitioning species is greater than it is in water because complementary charges have been preorganized and stabilized by the folding energy - which has been optimized by evolution to accommodate the mechanical and electronic strain that is produced as a consequence of this preorganization.

**THE PHENOTYPE OF STRAIN**

Strain, whether it is electronic or mechanical, is a phenotype of both the entatic and preorganization theories. Although the preorganization theory denies an evolutionary advantage in the direct use of steric strain in the enhancement of rates, it does not deny the evolution of steric strain as a consequence of the optimization of factors that do contribute directly to rate enhancing features (the electrostatic effect). For example, the optimization of protein dipole positions may require sterically straining the backbone or the side-chain of residues - a job that may be performed by the evolution of fold stabilizing features. Certainly, electronic strain is a factor in the preorganization of protein dipoles, especially in the case where the protein folds before the delivery of a functional countercharge. In support of both the entatic and the preorganization theories surveys of backbone conformations in protein structures have found hidden steric strain in protein molecules and an increased number of rare backbone conformations at functional sites (Herzberg & Moult 1991, Karplus 1996, Petock et al. 2003).

The forces that aid in the function and stability of enzymes are the same forces that aid in the function and stability of all proteins. Features that enhance binding and specificity (exposed hydrophobic patches, preoriented dipoles, specificity loops) gain function from folding energy. A growing number of studies show an increase in fold-stability upon the replacement of functional residues, giving credence to the notion that functional residues are strained to some degree and that the protein fold pays to accommodate this strain. These mutational studies suggest that function and stability are evolutionarily connected, antagonistic forces that `are corresponding in nature`. Stability-function studies done on barnase (Meiering et al. 1992), barstar (Schreiber et al. 1994), staphylococcal nuclease (Hibler et al. 1987, Poole et al. 1991), thioredoxin reductase (Gleason 1992), Cro repressor (Pakula & Sauer 1989), T4 lysozyme (Shoichet et al. 1995), AmpC-beta lactamase (Beadle & Shoichet 2002), and o-succinylbenzoate synthase (Nagatani et al. 2007), to name a few, show a relationship with decreased activity and increased stability. Some of these effects are quite large. Active site residue mutations of AmpC-beta lactamase caused decreases in activity by $10^3$ - $10^5$ compared to WT while concomitantly increasing the stability by up to 4.7 kcal/mol - a 30% increase in stability (Beadle & Shoichet 2002).

Although the results of these and other studies suggest that stability trade-offs for function may be a general property of biological catalysts, when analyzing these effects it is important to consider the type of function that the residue has. In general, alanine substitutions at functional sites render more stable proteins, but this is not always the case. The studies above address two different functions: binding (hydrophobic patches at the surface) and catalysis. In the case of binding, it is easy to see why alanine substitutions of the hydrophobic residues at the surface increase the stability of the overall protein, but in the case of catalysis it is not so cut and dry. For example, an alanine substitution at S64, an AmpC-beta lactamase residue believed to participate directly in catalysis, actually decreases the stability (Beadle & Shoichet 2002). The preorganization theory predicts that there are residues (or groups) that participate directly in catalysis and there are residues that are part of the preorganized network aimed at enhancing the function of residues that participate directly in catalysis. Thus, the stability of a fold after removal of a catalytic residue depends on how that residue functions in catalysis.

A good way to study the effect of residues that are part of the preorganized dipole setup is with metalloenzymes, where the functional charge is on the metal and the first-shell residues represent the dipole network aimed at the enhancement of the metal function. Work done by Nagatani et al., alanine substitutions at metal coordinating residues of *o*-succinylbenzoate synthase, a representative member of the enolase superfamily, significantly increases the stability of the enzyme (0.26- 4.23 kcal/mol) (Nagatani et al. 2007). This result indicates that the protein fold pays the cost of the electronic strain produced by the clustering of like charges required in forming the functional metal site.

## FOLD, FUNCTION, AND FAMILY

Studies show that there is low to no correlation between a particular fold and a particular function (Martin et al. 1998, Hegyi & Gerstein 1999). A comprehensive survey of the yeast genome found that some functions are associated with as many as seven folds and that it is not uncommon for a fold to have as many as sixteen functions (Hegyi & Gerstein 1999). Because the unsatisfied forces that destabilize proteins are the very forces that enhance catalytic function, pressure is put toward scaffolds that optimize the stability of the folded state. This may shed light on why an estimated ten percent of enzymes have the TIM Barrel fold. Analysis of non-homologous TIM Barrel structures show: most residues in the TIM Barrel fold are in favored conformations, helices are near ideal, and the beta barrel residues are almost completely shielded from solvent (Babu, TIM Barrel analysis, unpublished, Brandon & Tooze 1999). This design supports features that enhance the hydrophobic effect, and, concurrently, minimize charge destabilization. As far as enzymes go, TIM Barrels may represent structural perfection. But, it is important to realize that it may undermine the fitness of an organism if certain enzymes are too efficient at catalyzing their reactions. Less stable proteins, or proteins that are dynamic to some degree, may help regulate overzealous catalysis by lowering the availability of the functional space provided by the enzyme.

Enzymes that have conserved features that contribute to a particular function, although the overall chemical reaction catalyzed may be different, are part of the same mechanistically diverse superfamily (Gerlt & Babbitt 2001, Glasner et al. 2006, Glasner et al. 2007). A recent review by Glasner and colleagues discusses the mechanistic and structural aspects of four superfamilies (Glasner et al. 2006). Members within a family show conservation of active-site residues and a striking likeness in the positions of those residues. For example, the members that constitute the enolase superfamily share this function: they must remove a proton alpha to a carboxylate group and stabilize an enolate intermediate, which is a charged species. The crystal structure alignments reveal that the residues that coordinate the magnesium ion, which stabilizes the enolate intermediate, are virtually superimposable across all members of the family. In support of the function-stability trade-off hypothesis, alanine substitutions for the metal coordinating residues in *o*-succinylbenzoate synthase (OSBS), a representative member of the enolase family, stabilize the protein by up to 4 kcal/mol (Nagatani et al. 2007).

**EVOLVABILITY**

Recently, light has been shed on *how* a scaffold can accommodate a new function. Using phylogenetic, functional, and structural analysis the evolutionary trajectory of an ancient precursor to a modern protein shows that mutations that cause functional change *follow* stabilizing mutations that have no immediate consequence (Ortlund et al. 2007). This mechanism of evolvability allows the enzyme to maintain essential function and experiment with new ones.

The fitness of an organism may depend on whether or not an enzyme can discriminate between similar substrates. Directed evolution experiments that screen for the adaptation of functions that share the same chemistry but work on different substrates show that specificity evolves with only a few mutations - first to an intermediate enzyme, which may exhibit little or no loss in native activity, and then to an enzyme specific for the new substrate (reviewed by Aharoni et al. 2005, Khersonsky et al. 2006). Some studies show that the intermediate enzyme, dubbed promiscuous for not committing to a single substrate, has an open and/or more dynamic binding site. The surprising finding that broadening specificity can evolve with little or no decrease in the original function is reshaping the understanding of enzyme evolvability and the timing of gene duplication events. Gene duplication does ensure active, specific enzymes, but it need not happen before adaptation to a new function: the results of directed evolution experiments indicate that for new functions, that are similar to old functions, negative trade-offs tend to be weak.

Functional specificity can be acquired independently from functional catalysis. A system that has evolved to perform a particular chemistry may adapt a selection function to discriminate between similar substrates if the products of some of those substrates decrease fitness. Under selective pressure the selection function may breed complexity in terms of organized structure and architecture. Alternatively, an enzyme may broaden the substrate landscape by 'down-shifting' the selection function - loosening specificity architecture - a process that may not affect the catalysis function, which is paid for by complex interactions elsewhere.

**ON ENZYME EVOLUTION AND EMERGENCE**

Simply put, complex specifiable molecules exist because they have what it takes to get themselves replicated. Studies show that enzymes, and proteins in general, store functional energy in their folds. They are the manifestation of the coevolution of strength and strain. The organization of matter produced by information passed down from a very special feedback cycle. Functional features like preorganized-same-charge-dipoles and hydrophobic patches place selective pressure on scaffolds that can do the work necessary in setting up the functional site. In turn, complex structural motifs like alpha helices and beta sheets place selective pressure on the functions that create them. As this relationship between strength and strain endures, layers of organization emerge.

The reactions that produce larger molecules often require catalytic assistance. For example, making a carbon-carbon bond is not easy; the unique properties of carbon can result in high-energy charge-separated species that rarely manifest. It has been shown that metal sulfides have the capability to bind small carbon molecules, like carbon dioxide, and catalyze the formation of carbon-carbon bonds on their surface (Huber & Wachtershauser 1997, Cody et al 2000, Cody et al. 2001, Cody et al. 2004). But, carbon fixation alone does not explain the emergence of heredity - which could have begun as an autocatalytic system. It has been postulated (reviewed in Wachtershauser 2007) that carbon fixation products formed on metal sulfide surfaces may, in turn, bind to metal sulfide surfaces and enhance the rate of carbon-fixation chemistry – producing a feedback cycle, which may be at the origin of heredity. By chance, molecules could have been produced that had enough stored folding energy to do the work necessary to provide a space that would enhance the rate of their own production. This means that *by* existing the molecule increases its probability *of* existing, in other words, by existing, the molecule automatically manifests instructions on how to recreate itself. The ancestral replicators, born from nothing but the simple laws of chemistry, may have emerged from the same special relationship between strength and strain manifest in the macromolecules of today.